\documentclass[preprint,showpacs,preprintnumbers,amsmath,amssymb]{revtex4}

\usepackage{amsmath}
\usepackage{graphicx}
\usepackage{amssymb}
\usepackage{color}

\newcommand{\be}{\begin{eqnarray}}
\newcommand{\ee}{\end{eqnarray}}

\newcommand{\p}{\partial}

\newcommand{\nn}{\nonumber}

\newcommand{\diag}{\mathop{\rm diag}}

\newcommand{\cA}{{\mathcal A}}

\newcommand{\cF}{{\mathcal F}}
\newcommand{\cQ}{{\mathbf F}}

\newcommand{\cD}{{\mathcal D}}
\newcommand{\cG}{{\mathcal G}}
\newcommand{\fG}{{\mathbf G}}
\newcommand{\cW}{{\mathcal W}}

\newcommand{\dchi}{\delta \chi }

\newcommand{\1}{\mspace{1mu}}

\newcommand{\fGa}{\mathbf \Gamma}
\newcommand{\vGa}{\varGamma}

\newcommand{\cR}{\mathcal R}

\newcommand{\mM}{\mathsf M}
\newcommand{\mN}{\mathsf N}

\newcommand{\mA}{\mathsf A}

\newcommand{\fL}{\mathbf L}
\newcommand{\fM}{\mathbf M}
\newcommand{\fN}{\mathbf N}
\newcommand{\fP}{\mathbf P}
\newcommand{\fQ}{\mathbf Q}

\newcommand{\fA}{\mathbf A}
\newcommand{\fB}{\mathbf B}
\newcommand{\fC}{\mathbf C}
\newcommand{\fD}{\mathbf D}

\newcommand{\bet}{\bar{\eth}}

\newcommand{\vOm}{\varOmega }
\newcommand{\cH}{\mathbf A }

\newcommand{\chih}{\hat{\chi}}

\newcommand{\hx}{\hat{x}}

\newcommand{\tx}{\tilde{x}}
\newcommand{\mG}{\mathsf{G}}

\newcommand{\hPsi}{\hat{\Psi}}

\newcommand{\vPsi}{\varPsi}

\begin{document}
\title{Hyperunified field theory and \\ 
Taiji program in space for GWD }

\author{Yue-Liang Wu$^*$ }

\address{International Centre for Theoretical Physics Asia-Pacific (ICTP-AP)\\
Institute of Theoretical Physics, Chinese Academy of Sciences (ITP-CAS)\\
University of Chinese Academy of Sciences (UCAS) \\
$^*$E-mail: ylwu@itp.ac.cn, ylwu@ucas.ac.cn
}



\begin{abstract}
In this talk, I present the recently established hyperunified field theory (HUFT) \cite{YLWU1,YLWU2} for all basic forces and elementary particles within the framework of gravitational quantum field theory (GQFT)\cite{YLWU3,YLWU4} in hyper-spacetime. GQFT treats gravity as a gauge theory in the framework of quantum field theory to avoid the long term obstacle between the general relativity and quantum mechanics.  HUFT is built based on the guiding principle: the dimension of hyper-spacetime correlates to intrinsic quantum numbers of basic building blocks of nature, and the action describing the laws of nature obeys the gauge invariance and coordinate independence, which is more fundamental than that proposed by Einstein for general relativity.  The basic gravitational field is defined in biframe hyper-spacetime as a bicovariant vector field, it is a gauge-type hyper-gravifield rather than a metric field. HUFT is characterized by a bimaximal Poincar\'e and hyper-spin gauge symmetry PO(1,$D_h$-1)$\Join$SP(1,$D_h$-1) with a global and local conformal scaling invariance in biframe hyper-spacetime. The gravitational origin of gauge symmetry is revealed through the hyper-gravifield that plays an essential role as a Goldstone-like field, which enables us to demonstrate the gauge-gravity and gravity-geometry correspondences and to corroborate the gravitational gauge-geometry duality with an emergent hidden general linear group symmetry GL($D_h$,R). The Taiji Program in Space for the gravitational wave detection in China\cite{Nature,Taiji,HUWU} is briefly outlined.
\end{abstract}


\maketitle

\section{Introduction}

It has become a great attempt for many physicists to unify gravity described by the Einstein general theory of relativity (GR)\cite{GR} with other basic forces characterized by the Yang-Mills gauge theory\cite{YM}.  Recently, a hyperunified field theory (HUFT)\cite{YLWU1,YLWU2} has been established within the framework of gravitational quantum field theory(GQFT)\cite{YLWU3,YLWU4}. The framework of GQFT has been built based on the spin and scaling gauge invariances by treating the gravitational interaction on the same footing as the electroweak and strong interactions. GQFT enables us to make a unified description for the four basic forces within the framework of quantum field theory (QFT) in a globally flat Minkowski spacetime. GQFT is constructed with the postulates of gauge invariance and coordinate independence, which become more general and fundamental than the postulate of general covariance that requires the general linear group symmetry GL(4,R) in the curved Riemannian spacetime. In GQFT, all the basic forces are governed by gauge symmetries. Instead of the metric field in GR, a bicovariant vector field defined in a biframe spacetime is introduced as a basic gauge-type {\it gravifield} to characterize the gravitational interaction. The concept of {\it biframe spacetime} plays an essential role in GQFT.  One frame spacetime is considered as an inertial reference frame for describing the motions of basic fields, so that it enables us to derive the well-defined conservation laws and to make a physically meaningful definition for space and time with the standard ways proposed in special relativity (SR). Such a frame spacetime is known in SR as a globally flat Minkowski spacetime of the coordinate system. The other frame spacetime is a locally flat non-coordinate {\it gravifield spacetime} which is viewed as an intrinsic interaction frame for characterizing the dynamics of basic fields. 

When treating space and time equally as four dimensional spacetime in SR and applying to quantum mechanics (QM), Dirac obtained a relativistic spinor theory\cite{Dirac} which provided a successful unit of QM and SR.  It is such a Dirac spinor theory that reveals an interesting correlation between the dimensions of coordinate spacetime and the basic quantum numbers of the Dirac spinor. To see more explicitly for such a correlation, we have recently investigated a massless Dirac spinor, which is known to get new symmetries corresponding to the chirality spin and conformal scaling transformations. It has been shown that when treating the chirality spin on the same footing as the boost spin and helicity spin of the Dirac spinor, we arrived at a generalized Dirac spinor theory in six dimensional spacetime with the Lorentz and spin symmetries SO(1,5)$\cong$SP(1,5)\cite{YLWU5}. It indicates that the chirality spin of the Dirac spinor field does correlate to a rotation in the extra two spatial dimensions, which reflects, in the other way around, the correlation between the basic quantum numbers of Dirac spinor and the dimensions of coordinate spacetime as the Dirac spinor has actually eight real components

Inspired by the relativistic Dirac spinor theory for the massive Dirac spinor and the generalized Dirac spinor theory in six dimensional spacetime for the massless Dirac spinor\cite{YLWU5}, it is natural to make the hypotheses that all the spin-like charges of elementary particles should be treated on the same footing as a {\it hyper-spin charge} and the {\it hyper-spinor structures} of elementary particles are correlated with the geometric properties of high dimensional {\it hyper-spacetime}. With such hypotheses and the postulates in building GQFT\cite{YLWU3,YLWU4}, we arrived at the following general guiding principle for establishing a reliable HUFT\cite{YLWU1,YLWU2}:
 
(i) the dimension and structure of hyper-spacetime correlate to the intrinsic quantum numbers and independent degrees of freedom of basic building blocks of nature; 

(ii) the basic action characterizing the laws of nature obeys the gauge invariance and coordinate independence. 

\section{Unification of all basic forces and elementary particles }

It is known that the three basic forces in the standard model (SM)\cite{SM1,SM2,SM3,SM4,SM5} can be described by a minimal grand unified theory (GUT)\cite{GUT1,GUT2,GUT3} by unifying the electric charge, isospin charges and color-spin charges. In general,  when treating all the spin-like charges of basic building blocks on the same footing as a {\it hyper-spin charge}, all elementary particles should be merged into a column vector in a spinor representation of {\it hyper-spacetime}.  For all quarks and leptons as elementary particles in SM, they can be incorporated into a Majorana-type hyper-spinor field $ \varPsi(\hx)$ defined in a spinor representation of 18-dimensions,
\be \label{19DMHSF}
\varPsi(\hx) =  \binom{\hPsi(\hx) }{\hPsi^{'}(\hx) } =
\begin{pmatrix}
\Psi(\hx) \\
\Psi^{c}(\hx)\\
\Psi'(\hx) \\
-\Psi^{'c}(\hx)
\end{pmatrix}\, ,
\ee
where the spinors $\Psi(\hx)$ and $\Psi'(\hx)$ are Dirac-type hyper-spinor fields defined in the 14-dimensions with their charge conjugated ones $\Psi^{c}(\hx)$ and $\Psi^{'c}(\hx)$, they have the following structure, 
\be \label{SS1}
 \Psi  \equiv \Psi_{W} + \Psi_{E} \equiv \Psi_1 + i \Psi_2, \quad \Psi_{W, E} = \frac{1}{2} (1 \mp \gamma_{15}) \Psi , \quad \Psi_i \equiv  \Psi_{W i} + \Psi_{E i}. \nn
\ee
$\Psi_{W i}$ and $\Psi_{E i}$ ($i=1,2$) are a pair of {\it mirror hyper-spinor fields} called as {\it westward} and {\it eastward } hyper-spinor fields. Each {\it westward} spinor field $\Psi_{W i}$ represents a family of quarks and leptons in SM
\be
& & \Psi_{W\1 i}^{T} = [ (U_i^{r}, U_i^{b}, U_i^{g}, U_i^{w}, D^{r}_{ic}, D^{b}_{ic}, D^{g}_{ic}, D^{w}_{ic}, D_i^{r}, D_i^{b}, D_i^{g}, D_i^{w}, -(U^{r}_{ic}, U^{b}_{ic}, U^{g}_{ic}, U^{w}_{ic}))_L,  \nn \\
& & \; \ (U_i^{r}, U_i^{b}, U_i^{g}, U_i^{w}, D^{r}_{ic}, D^{b}_{ic}, D^{g}_{ic}, D^{w}_{ic}, D_i^{r},  D_i^{b}, D_i^{g}, D_i^{w}, -(U^{r}_{ic}, U^{b}_{ic}, U^{g}_{ic}, U^{w}_{ic}))_R]^T,
\ee
which is a Majorana-Weyl type hyper-spinor field in 14-dimensions with 64 independent degrees of freedom. $Q_i^{\alpha} \equiv (U_i^{\alpha}, D_i^{\alpha})$ are the Dirac spinors of quarks and leptons. The superscripts $\alpha= (r,\, b\, , g\, , w)$ represent the trichromatic colors (red, blue, green) for quarks and white color for leptons. $Q_{i c}^{\alpha} \equiv (U_{i c}^{\alpha},  D_{i c}^{\alpha})$ are the conjugated ones $Q_{i\1 c}^{\alpha} = C_4 \bar{Q}_i^T$ with $C_4$ defined in 4D-spacetime. The subscripts $``L"$ and $``R"$ in $Q_{i L,R}^{\alpha}$ label the left-handed and right-handed Dirac spinors.  The {\it westward} and {\it eastward } hyper-spinor fields $\Psi_{W i}$ or $\Psi_{E i}$ satisfy the Majorana-Weyl type condition defined in the 14-dimensions with a W-parity\cite{YLWU5},
\be
& & \Psi_{W i,E i}^{c} = C_{14} \bar{\Psi}_{W i, E i}^T = \gamma_5 \Psi_{W i, E i}, \quad \gamma_{15} \Psi_{W i, E i} = \mp \Psi_{W i, E i},  \nonumber \\
& & C_{14}  =  \sigma_1 \otimes \sigma_2 \otimes \sigma_2 \otimes \sigma_0 \otimes \sigma_0 \otimes \gamma_5 C_4, \quad  \gamma_{15} = \gamma_{11}\otimes\gamma_5  . \nonumber 
\ee

From the first guiding principle of HUFT, the dimension and structure of the hyper-spacetime is determined by the given spinor structure with a maximal symmetry. Let us first consider the freely moving massless Majorana-type hyper-spinor field $ \varPsi(\hx)$, its action in the irreducible spinor representation is found to be,
\be \label{Uaction1}
I_H = \int [d\hx] \, \frac{1}{2} \bar{\vPsi}(\hx) \varGamma^{\fA} \, \delta_{\fA}^{\;\; \fM} i \partial_{\fM} \vPsi(\hx) \, ,  
\ee
with the minimal dimension $ \fA\, , \fM =0,1,2,3,5, \cdots, D_h=19$. Where $\delta_{\fA}^{\;\;\fM} $ is the Kronecker symbol, $\partial_{\fM}= \partial/\partial x^{\fM}$  the partial derivative. The Latin alphabets $\fA, \fB\cdots$ and the Latin alphabets starting from $\fM, \fN$ are adopted to distinguish vector indices in non-coordinate spacetime and coordinate spacetime, respectively.  They are raised and lowered by the constant metric matrices, $\eta^{\fA\fB} $ or $\eta_{\fA\fB} =$ diag.$(1,-1,\cdots,-1)$, and $\eta^{\fM\fN} $ or $\eta_{\mM\mN} =$ diag.$(1,-1,\cdots,-1)$. $\Gamma^{\fA}$ is the vector $\gamma$-matrix defined in the spinor representation of 19-dimensional spacetime, its explicit form with the components $\varGamma^{\fA} \equiv ( \varGamma^a, \varGamma^{A}, \varGamma^{m} )$ ($a=0,1,2,3$, $A = 5,\cdots, 14$, $m = 15,\cdots, D_h$) is found to be 
\be \label{gamma}
& & \varGamma^{a} = \; \sigma_0 \otimes \sigma_0 \otimes I_{32}\otimes \gamma^a\, ,  \quad \varGamma^{A} = i \sigma_0 \otimes \sigma_0 \otimes \Gamma^{A}\otimes \gamma_5\, , \nn \\
& & \varGamma^{15} = i \sigma_2 \otimes \sigma_3\otimes \gamma^{11} \otimes \gamma_5\, ,\quad \varGamma^{16} = i \sigma_1 \otimes \sigma_0\otimes  \gamma^{11} \otimes \gamma_5\, , \nn \\
& & \varGamma^{17} = i \sigma_2 \otimes \sigma_1\otimes \gamma^{11} \otimes \gamma_5\, , \quad \varGamma^{18} = i \sigma_2 \otimes \sigma_2\otimes \gamma^{11}\otimes \gamma_5\, , \nn \\
& & \varGamma^{19} = \; \sigma_0 \otimes \sigma_0 \otimes I_{32}\otimes I_{4} \, , \nn
\ee
with $\sigma_i$ (i=1,2,3) the Pauli matrices, and $\sigma_0\equiv I_2$, $I_{4}$, $I_{32}$ the unit matrices. $\gamma^a$ and $\Gamma^A$ are $\gamma$-matrices defined in four and ten dimensions, respectively. The action Eq.(\ref{Uaction1}) possesses the global maximal hyper-spin and Lorentz symmetry 
\be
SP(1,D_h -1)\cong SO(1,D_h -1)\, , \quad D_h =19\, ,
\ee
with generators $\varSigma^{\fA\fB} = -  \varSigma^{\fB\fA} = i[\varGamma^{\fA}\, , \varGamma^{\fB}]/4$ and  $\varSigma^{\fA19} = - \varSigma^{19\fA} = i\varGamma^{\fA}/2\, $, $\fA, \fB = 0, 1,2,3,5,\cdots, 18$.

The action Eq.(\ref{Uaction1}) is also invariant under the translational transformation $x^{\fM}\to x^{\fM} + a^{\fM}$ and the discrete symmetries: charge conjugation $\mathcal{C}$, parity inversion $\mathcal{P}$ and time reversal $\mathcal{T}$  defined in 19-dimensional hyper-spacetime. The hyper-spinor field transforms as, 
\be
& & \mathcal{C} \vPsi(\hx ) \mathcal{C}^{-1}  \equiv \varPsi^{c}(\hx)  = C_{19}  \bar{\vPsi}^{T} ( \hx )= \vPsi(\hx)\, ,  \nn \\
& & \mathcal{P} \varPsi(\hx ) \mathcal{P}^{-1} = P_{19} \varPsi(\tx )\, , \quad \mathcal{T} \varPsi(\hx ) \mathcal{T}^{-1} = T_{19} \varPsi( -\tx ) \, 
\ee
with
\be
& & C_{19} = \varGamma_2\varGamma_0\varGamma_6\varGamma_8\varGamma_{10}\varGamma_{12}\varGamma_{14}  \varGamma_{16}\varGamma_{18} \, , \quad P_{19} = \varGamma^0, \nonumber \\
& & T_{19}= i \varGamma_1\varGamma_3 \varGamma_5\varGamma_7 \varGamma_9\varGamma_{11} \varGamma_{13}\varGamma_{15} \varGamma_{17}\gamma_{19}, \quad \tx = (x^0, -x^{1},\cdots, - x^{18}, x^{19}), \nonumber \\
& & C_{19}^{-1} \varGamma^{\fA} C_{19} = (\varGamma^{\fA})^{T}, \quad P_{19}^{-1} \varGamma^{\fA} P_{19} = ( \varGamma^{\mA})^{ \dagger},\quad T_{19}^{-1} \varGamma^{\fA} T_{19} = (\varGamma^{\fA})^{T}\, . \nn
\ee
The equation of motion for the freely moving hyper-spinor is obtained from Eq.(\ref{Uaction1}),
\be
 \varGamma^{\fA} \, \delta_{\fA}^{\;\; \fM} i \partial_{\fM} \vPsi(\hx) = 0\, , \quad \eta^{\fM\fN}\p_{\fM}\p_{\fN} \vPsi(\hx) = 0\, , \quad \eta^{\fM\fN} = \diag(1, -1, \cdots , -1), 
\ee
which leads to the generalized relativistic quantum theory for the hyper-spinor field in the globally flat Minkowski hyper-spacetime.

For the freely moving massless hyper-spinor field, the hyper-spin symmetry SP(1,$D_h$-1) and Lorentz symmetry SO(1,$D_h$-1) have to incorporate coherently. Based on the gauge principle, the fundamental interaction is governed by gauge symmetry. Let us take SP(1,$D_h$-1) as a hyper-spin gauge symmetry. It is natural to introduce the {\it hyper-spin gauge field} $\cA_{\fM}(\hx) \equiv \cA_{\fM}^{\; \fA\fB}(\hx)\, \frac{1}{2}\varSigma_{\fA\fB}$. By requiring the theory be invariant under both global Lorentz and local hyper-spin gauge transformations as well as global and local conformal scaling transformations, it is essential to initiate the {\it bicovaraint vector field} $\chih_{\fA}^{\;\, \fM}(\hx)$ and the {\it scaling scalar field} $\phi(\hx)$. 

The action Eq.(\ref{Uaction1}) shall be extended to be a gauge invariant one,
\be \label{Uaction2}
I_H =  \int [d\hx] \, \phi^{D_h-4} \chi(\hx) \1 \frac{1}{2} \bar{\vPsi}(\hx) \varGamma^{\fA} \, \chih_{\fA}^{\;\; \fM}(\hx) i \cD_{\fM} \vPsi(\hx) ,
\ee
with $i\cD_{\fM} \equiv i\partial_{\fM} + \cA_{\fM}$ a covariant derivative and $\chi(\hx)$ an inverse of determinant $\chih(\hx) = \det \chih_{\fA}^{\; \; \fM}(\hx)$. The above action Eq.(\ref{Uaction2}) possesses the bimaximal symmetry
\be \label{MS}
G = P(1,D_h\mbox{-}1)\times S(1) \Join SP(1,D_h\mbox{-}1)\times SG(1)\, ,
\ee
with P(1,$D_h$-1) = SO(1,$D_h$-1)$\ltimes P^{1,D_h\mbox{-}1}$ the Poincar\'e symmetry group in Minkowski hyper-spacetime, S(1)$\Join$SG(1) the global and local scaling symmetries.
 
A gravitational relativistic equation of motion is derived from the action Eq.(\ref{Uaction2}),
\be \label{EMSF}
& & \varGamma^{\fA} \hat{\chi}_{\fA}^{\;\; \fM}(\hx) i ( {\mathcal D} _{\fM} + {\mathsf V} _{\fM}(\hx) ) \vPsi(\hx) =   0 \, , 
\ee
its gauge invariant quadratic form is given by 
\be \label{EMSF2}
& & \chih^{\fM\fN} \hat{\nabla}_{\fM}  \hat{\cD} _{\fN}  \vPsi 
= \varSigma^{\fA\fB} \chih_{\fA}^{\;\; \fM} \chih_{\fB}^{\;\; \fN} [\, \cF_{\fM\fN} + i {\cal V}_{\fM\fN} - {\cal G}_{\fM\fN}^{\fA'} \chih_{\fA'}^{\;\; \fP} i \hat{\cD} _{\fP} \, ]  \vPsi , 
 \ee
with $\hat{\nabla}_{\fM}  \equiv \nabla_{\fM} +  {\mathsf V} _{\fM}$, $\hat{\cD}_{\fN} \equiv \cD_{\fN} + {\mathsf V} _{\fN}$,  $\hat{\nabla}_{\fM} \hat{\cD}_{\fN} \equiv \hat{\cD}_{\fM}\hat{\cD}_{\fN} + \fGa_{\fM\fN}^{\fP} \hat{\cD}_{\fP}$. 

$\cF_{\fM\fN}(\hx)$, ${\cal G}_{\fM\fN}^{\fA}$ and ${\cal V}_{\fM\fN}$  define the field strengths of the {\it hyper-spin gauge field} $\cA_{\fM}(\hx)$, the {\it hyper-gravifield} $\chi_{\fM}^{\; \fA}(\hx)$ and the induced-gauge field  ${\mathsf V}_{\fM}$, respectively, 
 \be \label{STGF}
 \cF_{\fM\fN} & = & \partial_{\fM} {\cal A}_{\fN}- \partial_{\fN} {\cal A}_{\fM} - i [{\cal A}_{\fM},  {\cal A}_{\fN} ] \equiv\cF_{\fM\fN}^{\fA\fB}\frac{1}{2}\varSigma_{\fA\fB}  \, , \nn \\
 {\cal G}_{\fM\fN}^{\fA} & = & \cD_{\fM} \chi_{\fN}^{\;\; \fA} - \cD_{\fN} \chi_{\fM}^{\;\; \fA};\quad  {\cal V}_{\fM\fN}  =   ( \partial_{\fM}  {\mathsf V} _{\fN} - \partial_{\fN}  {\mathsf V} _{\fM}), 
\ee
with $\chih^{\fM\fN} =  \chih_{\fA}^{\;\; \fM} \chih_{\fB}^{\;\; \fN} \eta^{\fA\fB}$, $\fGa_{\fM\fN}^{\fP}  =  \chih_{\fA}^{\;\; \fP}  \cD_{\fM}\chi_{\fN}^{\;\;\fA}$. Where $\chi_{\fM}^{\;\; \fA}(\hx)$ is a dual bicovaraint vector field and ${\mathsf V} _{\fM}$ is an {\it induced-gauge field}. They are defined as 
\be
& & \chi_{\fM}^{\;\; \fA} (\hx)\, \chih^{\;\; \fM}_{\fB}(\hx)   =  \eta^{\;\; \fA}_{\fB}\, , \;\, \chi_{\fM}^{\;\; \fA}(\hx) \chih^{\;\;\fN}_{\fA}(\hx) = \eta_{\fM}^{\;\;\fN}, \nonumber \\
& & {\mathsf V}_{\fM} (\hx) =  \frac{1}{2} \partial_{\fM}\ln (\chi \phi^{D_h-3})  -\frac{1}{2}\chih_{\fB}^{\;\; \fN}\cD_{\fN}\chi_{\fM}^{\;\; \fB} \, , \nn \\
& & \cD_{\fM}\chi_{\fN}^{\;\; \fA}  =  (\, \partial_{\fM} + \p_{\fM} \ln\phi\, ) \chi_{\fN}^{\;\; \fA} + \cA_{\fM\, \fB}^{\fA}  \chi_{\fN}^{\;\;\fB} \, .
\ee

The bicovariant vector field $\chi_{\fM}^{\;\; \fA}(\hx)$ is a gauge-type hyper-gravifield which characterizes the gravitational interaction in the hyper-spacetime.

\section{Hyperunified field theory within the framework of GQFT}

Based on the second guiding principle of HUFT, the basic action characterizing the laws of nature obeys the gauge invariance and coordinate independence,  let us define, from a dual basis $(\{\p_{\fM}\}, \{dx^{\fM}\})$ of a coordinate system, a non-coordinate basis  $\delta\chi^{\fA} \equiv \chi_{\fM}^{\;\; \fA} (\hx) dx^{\fM}$,  $\eth_{\fA} \equiv  \chih_{\fA}^{\;\, \fM}(\hx) \partial_{\fM}$ via the dual condition
\be \label{eth}
& &  \langle \delta \chi^{\fA}, \eth_{\fB}\rangle = \chi_{\fM}^{\;\; \fA}(\hx)  \chih_{\fB}^{\;\; \fN} (\hx)  \langle dx^{\fM} , \partial_{\fN} \rangle = \eta_{\fB}^{\;\, \fA}. 
\ee
Such a dual {\it hyper-gravifield basis} $(\{\eth_{\fA}\}, \{\delta \chi^{\fA}\})$ will span a non-coordinate locally flat {\it hyper-gravifield spacetime}.  

In such a {\it hyper-gravifield spacetime}, we can arrive at the gauge invariant and coordinate independent HUFT\cite{YLWU1,YLWU2},
\be
\label{Uaction3}
I_H & \equiv &  \int [\dchi]\, \phi^{D_h-4}\{\,  \frac{1}{2}\bar{\vPsi}\vGa^{\fC}i\cD_{\fC} \vPsi  - \frac{1}{4}\, [\, g_h^{-2}\tilde{\eta}^{\fC\fD\1 \fC'\fD'}_{\fA\fB\1 \fA'\fB' } \cF_{\fC\fD}^{\fA\fB}\cF_{\fC'\fD'}^{ \fA'\fB'} + {\cal W}_{\fC\fD} {\cal W}^{\fC\fD} \, ] \nn \\
& + & \alpha_E \phi^2 [ \frac{1}{4}  \, \tilde{\eta}^{\fC\fD\fC'\fD'}_{\fA\fA'} \cG_{\fC\fD}^{\fA}\cG_{\fC'\fD'}^{\fA'} - \eta_{\fA}^{\fC}\eta_{\fB}^{\fD}\cF_{\fC\fD}^{\fA\fB} ]  +  \frac{1}{2}\eta^{\fC\fD}\bet_{\fC} \phi \bet_{\fD}\phi  - \beta_E\1 \phi^4  \, \} ,
\ee
where $i\cD_{\fC} \equiv i \eth_{\fC}  + \cA_{\fC}$, $\bet_{\fC} \phi  \equiv (\eth_{\fC} - g_w W_{\fC}) \phi$, $\cA_{\fC}^{\fA\fB} \equiv \chih_{\fC}^{\;\, \fM} \cA_{\fM}^{\fA\fB}$ and $W_{\fC} \equiv \chih_{\fC}^{\;\, \fM} W_{\fM}$. $W_{\fM}$ as a Weyl gauge field\cite{Weyl} is introduced to characterize the conformal scaling gauge invariant dynamics of the {\it scaling scalar field} $\phi$. The corresponding field strengths are given as:  $\cF_{\fC\fD}^{\fA\fB} \equiv \cF_{\fM\fN}^{\fA\fB}  \chih^{\;\fM}_{\fC} \chih^{\;\fN}_{\fD} $, $\cG_{\fC\fD}^{\fA}\equiv \cG_{\fM\fN}^{\fA}\chih^{\;\fM}_{\fC} \chih^{\;\fN}_{\fD}$ and $\cW_{\fC\fD}  \equiv \cW_{\fM\fN}\chih^{\fM}_{\fC} \chih^{\fN}_{\fD}$ with $\cW_{\fM\fN} = \p_{\fM}W_{\fN} - \p_{\fN}W_{\fM}$. 

The constants $g_h$, $\alpha_E$ and $\beta_E$ are the couplings. $\tilde{\eta}^{\fC\fD\1 \fC'\fD'}_{ \fA\fB\1\fA'\fB'} $ is the tensor with a specific structure to keep the general conformal scaling gauge invariance\cite{YLWU1},
\be \label{tensor}
& & \tilde{\eta}^{\fC\fD\1 \fC'\fD'}_{ \fA\fB\1\fA'\fB'} \equiv \frac{1}{4}\1 \{[\eta^{\fC\fC'} \eta_{\fA\fA'} (\eta^{\fD\fD'} \eta_{\fB\fB'} - 2  \eta^{\fD}_{\fB'} \eta^{\fD'}_{\fB}) +  \eta^{(\fC,\fC'\leftrightarrow\fD, \fD' )} ]+  \eta_{(\fA,\fA'\leftrightarrow \fB,\fB' ) } \} \nn \\
& &\quad +   \frac{1}{4}\alpha_W\, \{\, [\, (\eta^{\fC}_{\fA'} \eta^{\fC'}_{\fA} - 2\eta^{\fC\fC'} \eta_{\fA\fA'}  )  \eta^{\fD}_{\fB'}  \eta^{\fD'}_{\fB}   +  \eta^{(\fC,\fC'\leftrightarrow\fD, \fD' )} \, ] +  \eta_{(\fA,\fA'\leftrightarrow \fB,\fB' ) }  \,  \}  \nn \\
& & \quad +  \frac{1}{2}\beta_W\, \{\, [\, (\eta_{\fA\fA'}  \eta^{\fC\fC'} - \eta^{\fC'}_{\fA}\eta^{\fC}_{\fA'}) \eta^{\fD}_{\fB} \eta^{\fD'}_{\fB'}+  \eta^{(\fC,\fC'\leftrightarrow\fD, \fD' )}\, ]+  \eta_{(\fA,\fA'\leftrightarrow \fB,\fB' )} \, \} , 
\ee
with $\alpha_W$ and $\beta_{W}$ the coupling constants. The tensor factor $\tilde{\eta}^{\fC\fD\fC'\fD'}_{\fA\fA'}$  takes a specific structure to ensure the general massless condition for the hyper-spin gauge field\cite{YLWU1},
\be  \label{tensorM}
 \tilde{\eta}^{\fC\fD\fC'\fD'}_{\fA\fA'}  & \equiv &    \eta^{\fC\fC'} \eta^{\fD\fD'} \eta_{\fA\fA'}  
+ [ \eta^{\fC\fC'} ( \eta_{\fA'}^{\fD} \eta_{\fA}^{\fD'}  -  2\eta_{\fA}^{\fD} \eta_{\fA'}^{\fD'}  ) +  \eta^{(\fC,\fC'\leftrightarrow\fD, \fD' )} ] . 
\ee

Taking the hyper-gravifield $\chi_{\fM}^{\;\fA}$ as a projection operator field defined in the biframe hyper-spacetime, it is not difficult to arrive at the gauge invariant action of HUFT\cite{YLWU1,YLWU2} within the framework of GQFT\cite{YLWU3,YLWU4},
\be \label{Uaction4}
I_H   &\equiv&  \int [d\hx]  \chi\, [ \frac{1}{2} \bar{\vPsi} \varGamma^{\fA} \chih_{\fA}^{\; \fM} i \cD_{\fM} \vPsi  - \frac{1}{4} (\tilde{\chi}^{\fM\fN\1 \fM'\fN'}_{\fA\fB\1 \fA'\fB' } \cF_{\fM\fN}^{\fA\fB}\cF_{\fM'\fN'}^{ \fA'\fB'} +  {\cal W}_{\fM\fN} {\cal W}^{\fM\fN} )  \nn \\
&  + & \alpha_E \phi^2 \frac{1}{4} \tilde{\chi}^{\fM\fN\fM'\fN'}_{\fA\fA'} \fG_{\fM\fN}^{\fA}\fG_{\fM'\fN'}^{\fA'}  +  \frac{1}{2}\hat{\chi}^{\fM\fN} d_{\fM} \phi d_{\fN}\phi  - \beta_E\phi^4 ] \phi^{D_h-4}  \nn \\
& + &  2\alpha_Eg_h\p_{\fM}(\chi \phi^{D_h-2} \cA_{\fN}^{\fN\fM}),
\ee
where we have introduced the following definitions
\be
& & \tilde{\chi}^{\fM\fN\1 \fM'\fN'}_{\fA\fB\1 \fA'\fB' }\equiv \chih_{\fC}^{\;\fM}\chih_{\fD}^{\;\fN} \chih_{\fC'}^{\;\fM'}\chih_{\fD'}^{\;\fN'} \tilde{\eta}^{\fC\fD\1 \fC'\fD'}_{\fA\fB\1 \fA'\fB' }, \;\; \tilde{\chi}^{\fM\fN\1 \fM'\fN'}_{\fA\fA'}\equiv \chih_{\fC}^{\;\fM}\chih_{\fD}^{\;\fN} \chih_{\fC'}^{\;\fM'}\chih_{\fD'}^{\;\fN'} \tilde{\eta}^{\fC\fD\1 \fC'\fD'}_{\fA\fA' }, \nn \\
& &  i \cD_{\fM} = i\p_{\fM} + g_h \cA_{\fM} , \qquad  d_{\fM} \phi =    (\p_{\fM} - g_wW_{\fM})\phi, \quad \cA_{\fN}^{\fN\fM} \equiv \chih_{\fA}^{\, \fN}\chih_{\fB}^{\, \fM}\cA_{\fN}^{\fA\fB} , \nn \\
& &  \fG_{\fM\fN}^{\fA} =  \hat{\p}_{\fM}\chi_{\fN}^{\; \fA}  -   \hat{\p}_{\fN}\chi_{\fM}^{\; \fA}; \;\; \hat{\p}_{\fM}\equiv \p_{\fM} + \p_{\fM} \ln\phi .
\ee
Note that a redefinition for the hyper-spin gauge field $\cA_{\fM}\to g_h\cA_{\fM}$ has been made. The last term in Eq.(\ref{Uaction4}) reflects a surface effect. 

In the basic action Eq.(\ref{Uaction4}), the hyper-spin gauge field $\cA_{\fM}^{\fA\fB}$ is massless, which is ensured by the tensor structure $\tilde{\eta}^{\fC\fD\fC'\fD'}_{\fA\fA'}$. The field strength $\fG_{\fM\fN}^{\fA}$ that describes the gauge gravitational interaction corroborates the {\it gauge-gravity correspondence}.

\section{Hidden symmetry and gravitational gauge-geometry duality}

It can be shown that the basic action Eq.(\ref{Uaction4}) gets an enlarged local symmetry,
\be
G_{MS} = GL(D_h, R) \Join \mbox{SP(1},D_h\mbox{-1)} \times \mbox{SG(1)}.
\ee
Here the general linear group GL($D_h$,R) is viewed as an emergent hidden symmetry, which lays the foundation of Einstein's GR in four dimensions. Such a hidden symmetry emerges due to the fact that all gauge field strengths in basic action Eq.(\ref{Uaction4}) are antisymmetry tensors in hyper-spacetime, so that the connection $\vGa_{\fM\fN}^{\fP}$ characterizing the local symmetry GL($D_h$,R) is canceled due to its symmetric property $\vGa_{\fM\fN}^{\fP} = \vGa_{\fN\fM}^{\fP}$. Such a feature is actually a natural consequence of the guiding principle of gauge invariance and coordinate independence. Thus it always allows us to choose a globally flat Minkowski hyper-spacetime instead of a curved Riemannian hyper-spacetime to be the base spacetime. The basic action Eq.(\ref{Uaction4}) is considered to have the maximal global and local symmetries presented in Eq.(\ref{MS}). 

It has been shown that the hyper-spin gauge symmetry has a gravitational origin\cite{YLWU1}, so that we are able to decompose both the hyper-spin gauge field and its field strength into two parts 
\be \label{GOGS}
& & \cA_{\fM}^{\fA\fB}\equiv \vOm_{\fM}^{\fA\fB} + \cH_{\fM}^{\fA\fB}, \qquad \cF_{\fM\fN}^{\fA\fB} = \cR_{\fM\fN}^{\fA\fB} + \cQ_{\fM\fN}^{\fA\fB}\, , \nn \\
& & \cR_{\fM\fN}^{\fA\fB} =  \partial_{\fM} \vOm_{\fN}^{\fA\fB} - \partial_{\fN} \vOm_{\fM}^{\fA\fB} + \vOm_{\fM \fC}^{\fA} \vOm_{\fN}^{\fC \fB} -  \vOm_{\fN \fC}^{\fA} \vOm_{\fM}^{\fC \fB} \nn \\  
& & \cQ_{\fM\fN}^{\fA\fB} =   \cD_{\fM} \cH_{\fN}^{\fA\fB} - \cD_{\fN} \cH_{\fM}^{\fA\fB} +  \cH_{\fM \fC}^{\fA} \cH_{\fN}^{\fC \fB} -  \cH_{\fN \fC}^{\fA} \cH_{\fM}^{\fC \fB} \nn \\
& & \mG_{\fM\fN}^{\fA} = \p_{\fM} \chi_{\fN}^{\; \fA} -  \p_{\fN} \chi_{\fM}^{\;\fA},\; \cD_{\fM} \cH_{\fN}^{\fA\fB}  =   \partial_{\fM}  \cH_{\fN}^{\fA\fB}  +  \vOm_{\fM \fC}^{\fA} \cH_{\fN}^{\fC \fB} - \vOm_{\fM \fC}^{\fB} \cH_{\fN}^{\fC \fA},  \nn \\
& & \vOm_{\fM}^{\fA\fB}  =  \frac{1}{2}[ \hat{\chi}^{\fA\fN} \mG_{\fM\fN}^{\fB} - \hat{\chi}^{\fB\fN} \mG_{\fM\fN}^{\fA} -  \hat{\chi}^{\fA\fP}  \hat{\chi}^{\fB\fQ}  \mG_{\fP\fQ}^{\fC} \chi_{\fM \fC } ], 
\ee 
where $\vOm_{\fM}^{\fA\fB}$ transforms as a gauge field in the adjoint representation of SP(1,$D_h$-1) when $\chi_{\fM}^{\; \fA}$ transforms as a vector, while $\cH_{\fM}^{\fA\fB}$ transforms homogeneously. $\vOm_{\fM}^{\fA\fB}$ is referred as {\it hyper-spin gravigauge field} and  $\cH_{\fM}^{\fA\fB}$ as {\it hyper-spin homogauge field}.

As $\vOm_{\fM}^{\fA\fB}$ is completely determined by the gauge-type hyper-gravifield $\chi_{\fM}^{\; \fA}$, it does show that the hyper-spin gauge symmetry has a {\it gravitational origin}.  In this sense, $\chi_{\fM}^{\; \fA}$ behaves as a Goldstone-like boson, which enables us to obtain the following general relations for the gauge field and the field strength,  
\be \label{relation}
& & \vGa_{\fM\fQ}^{\fP}\equiv \chih_{\fA}^{\;\fP}(\p_{\fM}\chi_{\fQ}^{\;\fA} + \vOm_{\fM\fB}^{\fA}\chi_{\fQ}^{\fB}), \quad \cF_{\fM\fN}^{\fA\fB} =  (\cR_{\fM\fN}^{\fP\fQ} +  \cQ_{\fM\fN}^{\fP\fQ} )\chi^{\;\fA}_{\fP} \chi^{\;\fB}_{\fQ} \, , 
\ee
where $\cR_{\fM\fN}^{\fP\fQ}\equiv\cR_{\fM\fN\fQ'}^{\fP}\chih^{\fQ'\fQ} $ defines the Riemann tensor corresponding to  the {\it hyper-spacetime gravigauge field} $\vGa_{\fM\fQ}^{\fP}$. $\cQ_{\fM\fN}^{\fP\fQ}$ is the field strength of {\it hyper-spacetime homogauge field} $\cH_{\fM}^{\fP\fQ} \equiv  \chih_{\fA}^{\fP} \chih_{\fB}^{\fQ} \cH_{\fM}^{\fA\fB}$. Their explicit forms are given by  
\be \label{HSGF}
& & \cR_{\fM\fN\fQ}^{\fP} = \partial_{\fM} \vGa_{\fN\fQ}^{\fP} - \partial_{\fN} \vGa_{\fM\fQ}^{\fP}  + \vGa_{\fM\fL}^{\fP} \vGa_{\fN\fQ}^{\fL}  - \vGa_{\fN\fL}^{\fP} \vGa_{\fM\fQ}^{\fL}, \nn \\
& & \vGa_{\fM\fQ}^{\fP} =  \frac{1}{2} \hat{\chi}^{\fP\fL} (\, \p_{\fM} \chi_{\fQ \fL} + \p_{\fQ} \chi_{\fM \fL} - \p_{\fL}\chi_{\fM\fQ} \, )\, , \nn \\
& & \cQ_{\fM\fN}^{\fP\fQ} =  \nabla_{\fM} \cH_{\fN}^{\fP\fQ} - \nabla_{\fN} \cH_{\fM}^{\fP\fQ} + \cH_{\fM\fL}^{\fP} \cH_{\fN}^{\fL\fQ} - \cH_{\fN\fL}^{\fP} \cH_{\fM}^{\fL\fQ} \nn \\
& & \nabla_{\fM} \cH_{\fN}^{\fP\fQ} = \p_{\fM} \cH_{\fN}^{\fP\fQ} + \vGa_{\fM\fL}^{\fP} \cH_{\fN}^{\fL\fQ} + \vGa_{\fM\fL}^{\fQ}  \cH_{\fN}^{\fP\fL}, 
 \ee

Geometrically, $\vGa_{\fM\fQ}^{\fP}$ is a Christoffel symbol characterized by the hyper-gravimetric field $\chi_{\fM\fN}= \chi_{\fM}^{\; \fA}\chi_{\fN}^{\; \fB} \eta_{\fA\fB}$, which enables us to rewrite the basic action Eq.(\ref{Uaction4}) into an equivalent action in a hidden gauge formalism\cite{YLWU1,YLWU2},
\be \label{Uaction5}
I_H & = &  \int d\hx \chi \phi^{D_h-4}  \{ \frac{1}{2}\bar{\vPsi}\vGa^{\fM} [ i \p_{\fM}  +  (\varXi_{\fM}^{\fP\fQ} + g_h\cH_{\fM}^{\fP\fQ} ) \frac{1}{2}\varSigma_{\fP\fQ} ] \vPsi  \nonumber \\
& - & \frac{1}{4} (\, \tilde{\chi}^{\fM\fN\1 \fM'\fN'}_{\fP\fQ\1\fP'\fQ'}  \cQ_{\fM\fN}^{\fP\fQ} \cQ_{\fM'\fN'}^{\fP'\fQ'}  + {\cal W}_{\fM\fN} {\cal W}^{\fM\fN} \, ) + \frac{1}{2}\hat{\chi}^{\fM\fN} d_{\fM} \phi d_{\fN}\phi - \beta_E\1 \phi^4   \nonumber \\
& + & \alpha_E  [ \,\phi^2\cR - (D_h-1)(D_h-2)\p_{\fM}\phi\p^{\fM}\phi  \, ]   \}  +  2\alpha_Eg_h\p_{\fM}(\chi \phi^{D_h-2} \cH_{\fN}^{\fN\fM}) ,
\ee
with $\vGa^{\fM}\equiv\chih_{\fA}^{\fM}\vGa^{\fA}$, $\varXi_{\fM}^{\fP\fQ}\equiv \frac{1}{2} ( \chih^{\;\fP}_{\fC} \p_{\fM}\chih^{\fQ\fC} - \chih^{\;\fQ}_{\fC} \p_{\fM}\chih^{\fP\fC} )$ and $\cR \equiv -\cR_{\fM\fN\fQ}^{\fP}\eta_{\fP}^{\fM}\chih^{\fN\fQ}$. Where $\varXi_{\fM}^{\fP\fQ}$ defines a pure gauge-type field and $\cR$ is the Ricci scalar tensor. The tensor field $\tilde{\chi}^{\fM\fN\1 \fM'\fN'}_{\fP\fQ\1\fP'\fQ'}$ is defined via Eq.(\ref{tensor}) as 
\be
\tilde{\chi}^{\fM\fN\1 \fM'\fN'}_{\fP\fQ\1\fP'\fQ'}  \equiv  \chi^{\fA}_{\fP} \chi^{\fB}_{\fQ} \chi^{\fA'}_{\fP'} \chi^{\fB'}_{\fQ'} \chih_{\fC}^{\fM} \chih_{\fD}^{\fN} \chih_{\fC'}^{\fM'}\chih_{\fD'}^{\fN'} \tilde{\eta}^{\fC\fD\1 \fC'\fD'}_{ \fA\fB\1\fA'\fB'}. 
\ee

It is the structure Eq.(\ref{tensor}) that eliminates high derivative terms of Riemann and Ricci tensors in the action Eq.(\ref{Uaction5}). The gravitational interaction is characterized by the conformal scaling gauge invariant Einstein-Hilbert action, which explicitly verifies the {\it gravity-geometry correspondence}.

The equivalence of the actions Eqs. (\ref{Uaction3}), (\ref{Uaction4}) and (\ref{Uaction5}) can be realized explicitly by choosing the gauge fixing of SP(1,$D_h$-1) to be the unitary gauge through making the hyper-gravifield symmetric $\chi_{\fM\fA} = \chi_{\fA\fM}$, so that $\chi_{\fM\fN} = (\chi_{\fM\fA})^2$. Such an equivalence reveals the gravitational {\it gauge-geometry duality}.

\section{Taiji program in space for gravitational universe}

The discovery of gravitational waves by the LIGO collaboration\cite{LIGO}  has caused a significant influence on the development of basic research sciences. Space-based gravitational wave detection (GWD) is believed to be the next interesting target for exploring the gravitational universe as it can reach a wider range of gravitational radiation sources than the ground-based GWD can. Since the LISA/eLISA strategic plan\cite{LISA} on space-based GWD made in the 1990's, Chinese scientists have pushed forward the proposal for space-based GWD in China in the 2000's and promoted the potential cooperation with LISA\cite{Taiji1,Taiji2}. In 2016, Chinese Academy of Sciences (CAS) initiated a strategic priority research program for the pre-study of space-based GWD referred as the ``Taiji Program in Space" for gravitational physics\cite{Nature,Taiji,HUWU}.

The gravitational waves can provide a new window to explore the evolution of early universe and the nature of gravity. Unlike the ground-based GWD, the space-based GWD demands quite different key technology although it adopts the same detecting principle of a Michelson interferometer. For instance, space-based GWD has to make use of the gravitational reference sensors and the low noise micro-thruster to implement drag-free performance, and also the non-contacting discharging of test-masses for free flying test mass. It also requires a high precision measurement, such as the picometer optical assemblies for the resolution ranging between the test masses and the spacecraft and the high stability monolithic precision ranging of test-mass and spacecraft. In addition, the high stability telescope, high accuracy phase-meter and frequency stabilization as well as precision attitude control  are all needed for the precision ranging from spacecraft to spacecraft. 

The space-based GWD orbits the sun in order to reach high thermomechanical stability. The Taiji Program in Space proposes to use a triangle of three spacecrafts in orbit around the Sun. Laser beams are sent both ways between each pair of spacecraft, the gravitational wave effects are measured by the differences in the phase changes between the transmitted and received laser beams at each spacecraft. 

The Taiji Program in Space is proposed to detect GWs with frequencies ranging from 0.1mHz to 1.0Hz with a higher sensitivity around 0.01Hz-1Hz. The purpose of the Taiji program in Space is to investigate the most challenging issues concerning the massive black holes and the relevant fundamental problems, which includes how the intermediate mass seed black holes were formed in early universe, how the seed black hole grows into a large or extreme-large black hole, whether the dark matter could form black hole, what is the nature of gravity.

The preliminary design for the Taiji mission is based on 3 million kilometers separations between the spacecraft, it is expected to be launched around 2033. Before that, Taiji pathfinder consisting of a pair of the spacecrafts is proposed around 2023 to test the relevant key technology.

\section{Conclusions}

I have reported the HUFT which unifies all known basic forces and elementary particles based on a more fundamental principle. The unification of all quarks and leptons in SM into a single hyper-spinor field leads to a minimal 19-dimensional biframe hyper-spacetime with the bimaximal Poincar\'e and hyper-spin gauge symmetry PO(1,18)$\Join$SP(1,18) and the conformal scaling symmetry. The HUFT predicts the existence of mirror quarks and leptons and also vector-like hyper-spinors. As the HUFT is established from the bottom-up approach, it will be reasonable to figure out a realistic model to reproduce the SM with three families of quarks and leptons and to explain the matter-antimatter asymmetry and the dark matter component in the present universe. An appropriate symmetry breaking mechanism and a reliable dimension reduction are needed to reach four dimensional spacetime of the real world, so that HUFT is applicable to reveal the inflationary period of early universe and understand the observed accelerating expansion of present universe with the dominant dark energy component. The Taiji program in space is expected to probe the nature of gravity and provide a possible test on HUFT.

\centerline{{\bf Acknowledgement}}

The author is grateful to Harald Fritzsch for his invitation to deliver the talk at the conference on ``Particles and Cosmology". He would like to thank  K.K. Phua for his kind hospitality at NTU, Singapore. This work was supported in part by the National Science Foundation China (NSFC) under Grants No. 11690022 \& 11475237, and by the Strategic Priority Research Program of CAS under Grant No. XDB23030100, and by the CAS Center for Excellence in Particle Physics.

\bibliographystyle{ws-procs961x669}
\bibliography{ws-pro-sample}

\begin{thebibliography}{10}

\bibitem{YLWU1}
Y. L. Wu, ``Hyperunified field theory and gravitational gauge-geometry duality", Eur. Phys. J. {\bfseries C78}, 28 (2018); arXiv.1712.04537, 2018.
\bibitem{YLWU2} Y. L. Wu, ``Unified field theory of basic forces and elementary particles with gravitational origin of gauge symmetry", Sci. Bull. {\bf 62}, 1109 (2017); arXiv: 1705:06365.

\bibitem{YLWU3} Y. L. Wu, ``Quantum field theory of gravity with spin and scaling gauge invariance and spacetime dynamics with quantum inflation", Phys. Rev. {\bf D 93}, 024012 (2016).


\bibitem{YLWU4}
Y. L. Wu, ``Theory of quantum gravity beyond Einstein and space-time dynamics with quantum inflation", 
 Int. J. Mod. Phys. {\bfseries A30}, 1545002 (2015); arXiv: 1510.04720.

\bibitem{Nature} David Cyranoski, ``Chinese gravitational-wave hunt hits crunch time", Nature, 531, 150 (2016).

\bibitem{Taiji} Y. L. Wu, ``Taiji program in space and unified field theory in hyper-spacetime", talk presented at the International Symposium on Gravitational Wave (ISGW2017), University of Chinese Academy of Sciences (UCAS), Beijing, May 25-29, 2017.

\bibitem{HUWU} W.R. Hu and Y. L. Wu, ``Taiji Program in Space for gravitational wave physics and the nature of gravity", 
Natl. Sci. Rev. {\bf 4}, 685 (2017).

\bibitem{GR} A. Einstein, Sitz. Konigl. Preuss. Akad. Wiss., {\bf 25}, 844 (1915);  Annalen der Physik (ser. 4), 49, 769 (1916).


\bibitem{YM} C.N. Yang and R.L. Mills, Phys. Rev. {\bf 96}, 191 (1954).

\bibitem{Dirac} P.A. Dirac,  Proceedings of the Royal Society A: Math., Phys. and Engineering Sciences, 117 (778): 610 (1928). 

\bibitem{YLWU5} Y. L. Wu, ``Maximal symmetry and mass generation of Dirac fermions and gravitational gauge field theory in six-dimensional spacetime", Chinese Phys. {\bf C41}, 103106 (2017); arXiv: 1703:05436.

\bibitem{SM1}  S.L. Glashow, Nucl. Phys. {\bf 22}, 579 (1961).
\bibitem{SM2} S. Weinberg, Phys. Rev. Lett. {\bf 19}, 1264 (1967).
\bibitem{SM3} A. Salam, in Proceedings of the Eight Nobel Symposium, Stochholm, Sweden, 1968,
edited by N. Svartholm (Almqvist and Wikell, Stockholm, 1968).
\bibitem{SM4} D.~J.~Gross and F.~Wilczek, Phys.~Rev.~Lett.~{\bf 30}, 1343 (1973).
 \bibitem{SM5}  H.~D.~Politzer,
  Phys.~Rev.~Lett.~{\bf 30}, 1346 (1973).
  
\bibitem{GUT1} H. Georgi and S.L. Glashow,  Phys. Rev. Lett. 32, 438 (1974).
\bibitem{GUT2} H. Georgi, in Partic. and Fields 1974, ed. C. Carlson (Amer. Inst. of Phys., NY,
1975).
\bibitem{GUT3} H. Fritzsch and P. Minkowski,  Annals of Phys. 93, 193 (1975).
  
\bibitem{Weyl} H. Weyl, Sitz. Konigl. Preuss. Akad Wiss., {\bf 26}, 465 (1918).


\bibitem{LIGO} LIGO and Virgo Collaborations, Phys. Rev. Lett. {\bf 116}, 061102  (2016).

\bibitem{LISA} European Space Agency. Assessment Study Report. NGO: Revealing a hidden Universe: opening
a new chapter of discovery, ESA/SRE(2011)19, December 2011.
\bibitem{Taiji1} Y. L. Wu, ``Space Gravitational Wave Detection in China", talk presented at the First eLISA Consortium Meeting APC-Paris, France, Oct.22-23, 2012.
\bibitem{Taiji2} Y. L. Wu,  ``Introduction to gravitational wave detection in space at CAS China", talk presented at the Sino-German Symposium, Hannover, Germany, Sept. 14-17, 2015.



\end{thebibliography}


\end{document}